# Disentangling population strategies of two cladocerans adapted to different ultraviolet regimes

Carla E. Fernández[1,2]  | Melina Campero[1]  | Cintia Uvo[2]  | Lars-Anders Hansson[3]

[1]Unidad de Limnología y Recursos Acuáticos, Universidad Mayor de San Simón, Cochabamba, Bolivia

[2]Division of Water Resources Engineering, Lund University, Lund, Sweden

[3]Department of Biology/Aquatic Ecology, Lund University, Lund, Sweden

**Correspondence**
Carla E. Fernández, Unidad de Limnología y Recursos Acuáticos, Universidad Mayor de San Simón, Cochabamba, Bolivia.
Email: carlaelo.fernandez@gmail.com

**Funding information**
Swedish International Development Cooperation Agency (SIDA); GIRH project; Swedish Research Council

**Abstract**

Zooplankton have evolved several mechanisms to deal with environmental threats, such as ultraviolet radiation (UVR), and in order to identify strategies inherent to organisms exposed to different UVR environments, we here examine life-history traits of two lineages of *Daphnia pulex*. The lineages differed in the UVR dose they had received at their place of origin from extremely high UVR stress at high-altitude Bolivian lakes to low UVR stress near the sea level in temperate Sweden. Nine life-history variables of each lineage were analyzed in laboratory experiments in the presence and the absence of sub-lethal doses of UVR (UV-A band), and we identified trade-offs among variables through structural equation modeling (SEM). The UVR treatment was detrimental to almost all life-history variables of both lineages; however, the *Daphnia* historically exposed to higher doses of UVR (HighUV) showed a higher overall fecundity than those historically exposed to lower doses of UVR (LowUV). The total offspring and ephippia production, as well as the number of clutches and number of offspring at first reproduction, was directly affected by UVR in both lineages. Main differences between lineages involved indirect effects that affected offspring production as the age at first reproduction. We here show that organisms within the same species have developed different strategies as responses to UVR, although no increased physiological tolerance or plasticity was shown by the HighUV lineage. In addition to known tolerance strategies to UVR, including avoidance, prevention, or repairing of damages, we here propose a population strategy that includes early reproduction and high fertility, which we show compensated for the fitness loss imposed by UVR stress.

**KEYWORDS**

cladocera, *Daphnia*, life-history traits, stress, structural equation modeling, ultraviolet radiation

## 1 | INTRODUCTION

Natural climate differences across geographical regions affect the way species respond to alterations in environmental variables, including ultraviolet radiation (UVR). The interactions between climate change and UVR have been amply reported (Bornman et al., 2015; Erickson, Sulzberger, Zepp, & Austin, 2015; Williamson et al., 2014). It has also been suggested that surface levels of UVR will follow the recovery of stratospheric ozone during the first half of the twenty-first century, although changes in cloudiness and greenhouse gas emissions will play a key role (Medina-Sánchez et al., 2013). As UVR wavelengths may strongly affect vital compounds (such as DNA and proteins, leading to molecular damages, Sommaruga & Buma, 2000), different aspects of climate change interacting with UVR have been shown to affect

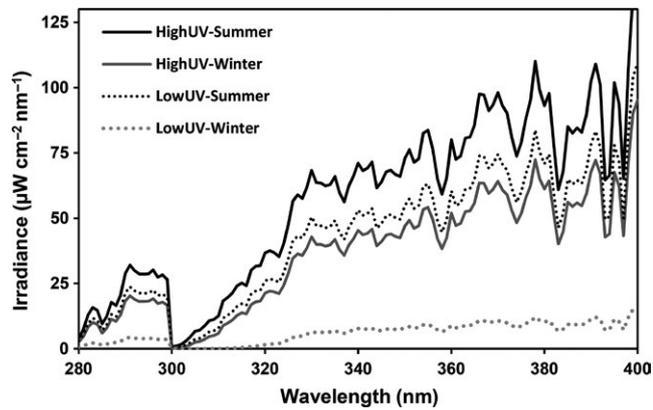

**FIGURE 1** Spectral simulated irradiances in summer and winter solstices of the high (Bolivia) and low (Sweden) UVR sites, respectively. Solid lines represent irradiances at the HighUV site, and dashed lines irradiances at the LowUV site (PV Lighthouse Pty, 2014)

plankton communities and their ecosystem functions (Häder et al., 2015; Hansson & Hylander, 2009; Hays, Richardson, & Robinson, 2005; Winder & Schindler, 2004). Some of them are linked to the colored dissolved organic matter (CDOM), which absorbs sunlight, including UVR, protecting plankton from the damaging UVR (Finkel et al., 2009). Reduced acidification, changes in runoff and land use, as well as bleaching of CDOM, may enhance the exposure of plankton communities to solar UVR (Erickson et al., 2015).

In lentic systems, zooplankton play a key role transferring energy and matter from primary producers to higher trophic levels. Thus, any factor that potentially damages zooplankton may affect the whole ecosystem via trophic interactions (Häder, Helbling, Williamson, & Worrest, 2011; Häder et al., 2015; Rautio & Tartarotti, 2010). Additionally, it has been suggested that interactions between species with different tolerance to UVR may also alter the dynamics and structure of the ecosystem (Fernández & Rejas, 2017; Hansson, 2004; Sommaruga, 2003). Two groups of strategies could influence the zooplankton tolerance to UVR: The first involves avoiding or preventing the damage, either behaviorally (through vertical migration, Hylander & Hansson, 2010; Kessler, Lockwood, Williamson, & Saros, 2008) or physiologically (through photo-protective compounds as melanin or carotenes, Rautio, Bonilla, & Vincent, 2009; Rautio & Korhola, 2002). The second group of strategies repairs the damage after it has happened, either by photo-enzymatic repair (PER; "light repair") or nucleotide excision repair (NER; "dark repair"; Hansson & Hylander, 2009; Mitchell & Karentz, 1993; Zagarese, Feldman, & Williamson, 1997).

As the use of any (or both) of the strategies is potentially costly (Gabriel, Luttbeg, Sih, & Tollrian, 2005; Tollrian & Heibl, 2004), it may be expected that lineages exposed to higher doses of UVR are adapted, or at least have a higher capacity to acclimatize to it. Herby, we refer to acclimatization as short-term physiological adjustments that occur during an organism's lifetime in response to transitory changes in environmental conditions. In contrast, adaptation is the process of genetic change that accumulates over a time scale of many generations in response to a specific environment of an organism (Morgan-Kiss, Priscu, Pocock, Gudynaite-Savitch, & Huner, 2006). This can be achieved by (1) adjusting to the environmental conditions through phenotypic plasticity, or (2) natural selection for more UVR resistant individuals in the population.

Although plasticity in a trait has been suggested as an adaptive mechanism that allows organism to optimally respond to environmental heterogeneity (Alpert & Simms, 2002; Callahan, Dhanoolal, & Ungerer, 2005), the limits between plasticity and adaptation are still poorly understood among zooplankton taxa. A geographical genetic differentiation among European *Daphnia magna* populations has been reported as evidence of local adaptation (Mitchell & Lampert, 2000). Hence, lineages with contrasting UVR regimens may display different traits to deal with UVR, and the way in which organisms handle the associated trade-offs between costs and benefits (e.g., life span vs. reproduction) of the displayed strategy should be reflected in their life-history responses.

Based on the assumption that actual life-histories are evolutionary stable while the trade-offs can change and evolve (Leroi, Chippindale, & Rose, 1994), we hypothesized that a lineage that has evolved in a high UVR environment would be less influenced by UVR stress than a lineage that has evolved under lower UVR exposure. Hence, we have here addressed the life-history patterns by which two nonpigmented lineages of *Daphnia pulex* from different UVR environments endure UVR exposure. Using structural equation modeling (SEM), we analyzed causal relationships in life-history traits of the following: (1) a lineage from a high UVR regime region in the high-altitude Andes (Bolivia), naturally exposed to extreme UVR conditions year around, and (2) a lineage from a low UVR regime region (Sweden), seasonally exposed to medium levels of UVR.

## 2 | MATERIALS AND METHODS

### 2.1 | Origin of lineages

Two lineages of *D. pulex* from different UVR environments were cultured under laboratory conditions. The first population (hereafter referred as HighUV) was hatched from ephippia originating from Totora-Khocha Lake (17.46°S–65.63°W), a high-altitude lake located at 3730 m a.s.l. in the Central Andes of Bolivia. This is one of many fishless lakes of glacial origin of the tropical Andes, which later became a reservoir. Due to their latitude, tropical high-altitude Andean lakes neither present ice cover during the winter nor stratification during the summer, but have extreme diel variations in temperature, strong winds, and extremely high levels of UVR (Aguilera, Lazzaro, & Coronel, 2013; Campero, Moreira, Lucano, & Rejas, 2011). The mean and maximum depths of the Totora-Khocha Lake are approximately 2.8 and 13 m, respectively, although depth can decrease dramatically during the dry season (April–November). Attenuation depths ($Z_{1\%}$ UVA) range from 0.40 to 2.98 m in the lakes of the area (Aguilera et al., 2013), depending on the CDOM content.

The second *Daphnia* lineage (hereafter referred as LowUV) was obtained from Lake Dalby Quarry, a low-altitude, low-UVR site, located in southern Sweden (55.66°N–13.40°E). The lake that is located at approximately 95 m a.s.l. has a maximum depth of about 10 m and contains a sparse fish population of rainbow trout (*Oncorhynchus mykiss*; Ekvall, Hylander, Walles, Yang, & Hansson, 2015).

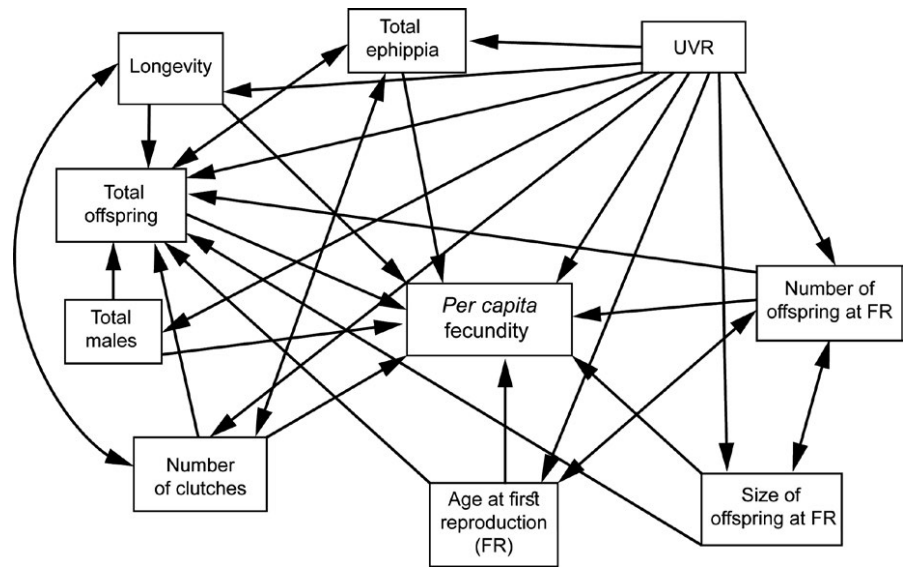

**FIGURE 2** Graphical representation of the a priori model used in the path analyses for the two *Daphnia pulex* lineages, including all variables and the potential relationships among them. One-headed arrows represent direct effects of one variable on another. Double-headed arrows represent covariance between variables

The calculated noon surface irradiances at 340 nm (UVA) for the Totora-Khocha Lake during the summer and winter solstices are approximately 71 and 45 µW/cm$^2$, respectively (PV Lighthouse Pty, 2014), whereas the irradiances at Dalby Quarry Lake are about 53 and 8 µW/cm$^2$ for summer and winter solstices, respectively (PV Lighthouse Pty, 2014). Thus, the HighUV lineage potentially receives an irradiance one-third higher in summer and almost five times higher in winter than the LowUV lineage (Figure 1). In addition, the LowUV lineage was kept in the laboratory for more than 100 generations exposed only to photosynthetically active radiation (PAR) provided by warm-white lamps (Aura Ultimate Long Life 36W) at 12:12 light:dark cycles.

## 2.2 | Experimental setup

In the laboratory, the acclimatization of both lineages was performed by keeping them separately at 20°C and 12:12 hr light:dark photoperiod (Taghavi, Farhadian, Soofiani, & Keivany, 2013), for 60 days. All animals were fed *ad libitum* three times a week with an algal culture mainly composed of *Scenedesmus* sp.

A life table experiment was performed to quantify life-history traits of both *Daphnia* lineages to UVR. Ten randomly chosen neonates of each lineage (~ 2 day old) were isolated and placed under UVR or Non-UVR treatments. UVR treatment was applied by covering the experimental flasks with a UVR-transparent acrylic sheet (UV-transmitting PLEXIGLAS® GS; Röhm GS 2458; Darmstadt, Germany) that has an average transmittance of 85% between 300 and 400 nm, while the Non-UVR treatment was achieved through a UVR-screening acrylic sheet (UV-absorbing PLEXIGLAS® GS; Röhm GS 233; Darmstadt, Germany) that cuts 100% radiation below 370 nm. For a full spectral transmittance of plexiglasses see Hansson, Hylander, and Sommaruga (2007). For both treatments, ultraviolet radiation at an intensity of 135 µW cm$^2$ was provided by three UVR fluorescent lamps (UVA-340; Q-panel) with a maximum emission in the UV-A band (340 nm). The illumination (30.3 µmol m$^2$/s intensity of photosynthetically active radiation (PAR)) was provided by four cool white fluorescent lamps (Aura Ultimate Life 36W). The daily provided UVR dose was 64.8 KJ/m$^2$, which represents approximately 20% and 7%, respectively, of the autumn mean daily doses in the UV-A band observed in Sweden (Danilov & Ekelund, 2001) and Bolivia (Villafañe, Andrade, Lairanat, Zaratti, & Helbling, 1999), respectively. Total irradiation measurements were made using a radiometer (IL 1400A; International Light; Newburyport, MA, USA) equipped with broadband sensors for UV-A (320–400 nm) and PAR (400–750).

The first generation of individuals was discarded to minimize interference from maternal effects (Lampert, 1993). Offspring born from the second clutch of synchronized mothers were used to start the experiment. Twenty replicate glass flasks with one *Daphnia* and 100 ml of dechlorinated tap water were exposed to each treatment in a 2 × 2 factorial design (lineage: HighUV/LowUV, UVA: presence/absence). Every other day 100 µl of an algal culture (529 µg/L ± 1.53; mean ± *SD*) mainly composed of *Scenedesmus* sp. was added to each flask. Animals were pipetted out to clean flasks and new medium once a week.

Flasks were checked every day, and offspring present were counted, measured (at first clutch), sexed, and discarded. The experiment was continued until all experimental individuals had died (approximately 100 days). Recorded life-history variables were the following: age at first reproduction (FR), size and number of neonates at FR, number of clutches, total number of born males, total offspring, total ephippia (dormant embryos) per female, and longevity. *Per capita* fecundity, that is, the number of offspring produced during the life span of each individual was used as a surrogate measure of the long-term individual contribution to population growth (Brommer, Gustafsson, Pietiäinen, & Merilä, 2004).

## 2.3 | Statistical analyses

Lineage and UVR effects on each variable were analyzed with a two-way ANOVA. Lineage and UVR exposure were used as independent variables, while each life-history trait was used as a dependent variable. Longevity was used as covariable in the analysis of the number of clutches, total number of born males, total offspring, and total ephippia per female.

Structural equations modeling is conceived to represent causal relationships among variables, which may have reciprocal influence between them either directly or through intermediary variables. To determine UVR-driven effects and causal relationships among life-history responses of both *Daphnia* lineages, we constructed an a priori theoretical model. In this model (Figure 2), we hypothesized that (1) all life-history variables are affected by UVR, (2) there are negative correlations between the numbers of ephippia and offspring produced and also between the number of clutches and the total ephippia per female, as time and energy applied to produce ephippia are higher than the ones for producing parthenogenetic offspring (Lynch, 1983), (3) the total offspring is affected by longevity, age at FR, number and size of offspring at FR, and number of clutches, (4) the age at FR is correlated to the number of offspring at FR, (5) the number of clutches is correlated to the longevity, (6) the number and size of offspring are mutually correlated, and finally, (7) that the *per capita* fecundity is affected by all life-history variables.

For model testing, all hypothesized paths were translated from the path diagram into a set of linear equations, one for each dependent variable. Lineage was treated as a grouping factor in a multigroup approach (Pugesek, Tomer, & Von Eye, 2003). UVR and Non-UVR treatments were incorporated into the model by presence/absence, assigning 1 to UVR exposed individuals and 0 to nonexposed ones. Due to differences in the scale of units between variables, all variables were standardized to have a mean of 0 and a standard deviation of 1.

We optimized the model by sequentially excluding nonsignificant paths ($p > .05$) from the models based on Akaike Information Criteria (AIC) scores. Comparative Fit Index [CFI; (Bentler, 1990)] and Tucker-Lewis Index [TLI; (Tucker & Lewis, 1973)] were also taken into account during the optimization process. These indices provide information on the model fit based on the discrepancy between the data and the hypothesized model, and a value close to 1 indicates a good fit. All paths that were not significant but whose exclusion would have worsened the model were retained in the model. All statistical analyses and figures were performed in R v3.2.5 (R Core Team, 2016) packages ggplot2 (Wickham, 2009) and Lavaan (Rosseel, 2012).

## 3 | RESULTS

### 3.1 | *Per capita* fecundity

The area below the fecundity curves for UVR treatments decreased compared to controls for both lineages, indicating a reduction in the *per capita* fecundity under UVR stress (Figure 3). The daily fecundity declined dramatically when the animals reached an age of 60–65 days. HighUV animals exposed to UVR showed

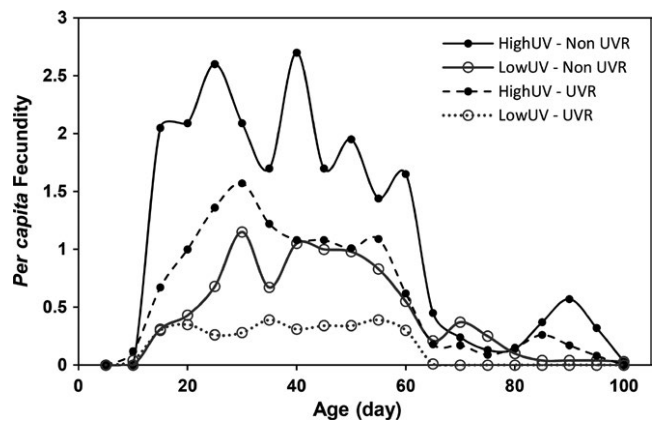

**FIGURE 3** Five-day average lifetime fecundity of HighUV and LowUV lineages, respectively, of *Daphnia pulex*. Non-UVR treatments are represented by solid lines, while dashed lines represent UVR treatments. Closed and open circles represent HighUV and LowUV lineages, respectively. *n* = 20 per treatment

maximum *per capita* fecundity simultaneously to the LowUV animals of the Non-UVR treatment (day 30), but at a higher level (Figure 3). The HighUV lineage extended reproduction to higher ages than LowUV animals, which stopped to reproduce at an age of 65 and 85 days with and without UVR, respectively. HighUV animals showed a higher overall fecundity than those of the LowUV lineage (Figure 3).

### 3.2 | ANOVA analysis

The UVR treatment was detrimental to almost all studied variables of both *Daphnia* lineages (Figure 4 and Table 1), as suggested by the reduced mean *per capita* fecundity (Figure 4a) and total offspring (Figure 4c). Average longevity ranged from 52 to 63 (maximum 100) days, and there were no differences between linages, nor between UVR treatments. Under the Non-UVR treatment, the LowUV lineage produced ephippia, and although both increased their ephippia production under UVR exposure ($p = .002$; Table 1), Tukey's post hoc test showed differences between lineages, but not between treatments (Figure 4d).

There was no production of males by the HighUV lineage in the absence of UVR, whereas there was a tendency to produce more males by the LowUV lineage when exposed to UVR ($p = .056$; Table 1; Figure 4e).

Both lineages showed reduced number of clutches in the UVR treatment, but this reduction was more pronounced for the HighUV lineage (Table 1; Figure 4f). The LowUV lineage animals at the Non-UVR treatment were, on average, more than 4 days older at first reproduction (FR) than the animals of the HighUV lineage, but at the UVR treatment, the first reproduction started at the same time as for the HighUV lineage (Figure 4g).

The number of neonates produced at first reproduction in the Non-UVR treatment was higher for the HighUV than the LowUV lineage (mean of 6.7 and 3.6, respectively), whereas both showed an average clutch size between 2 and 4 neonates in the UVR treatment

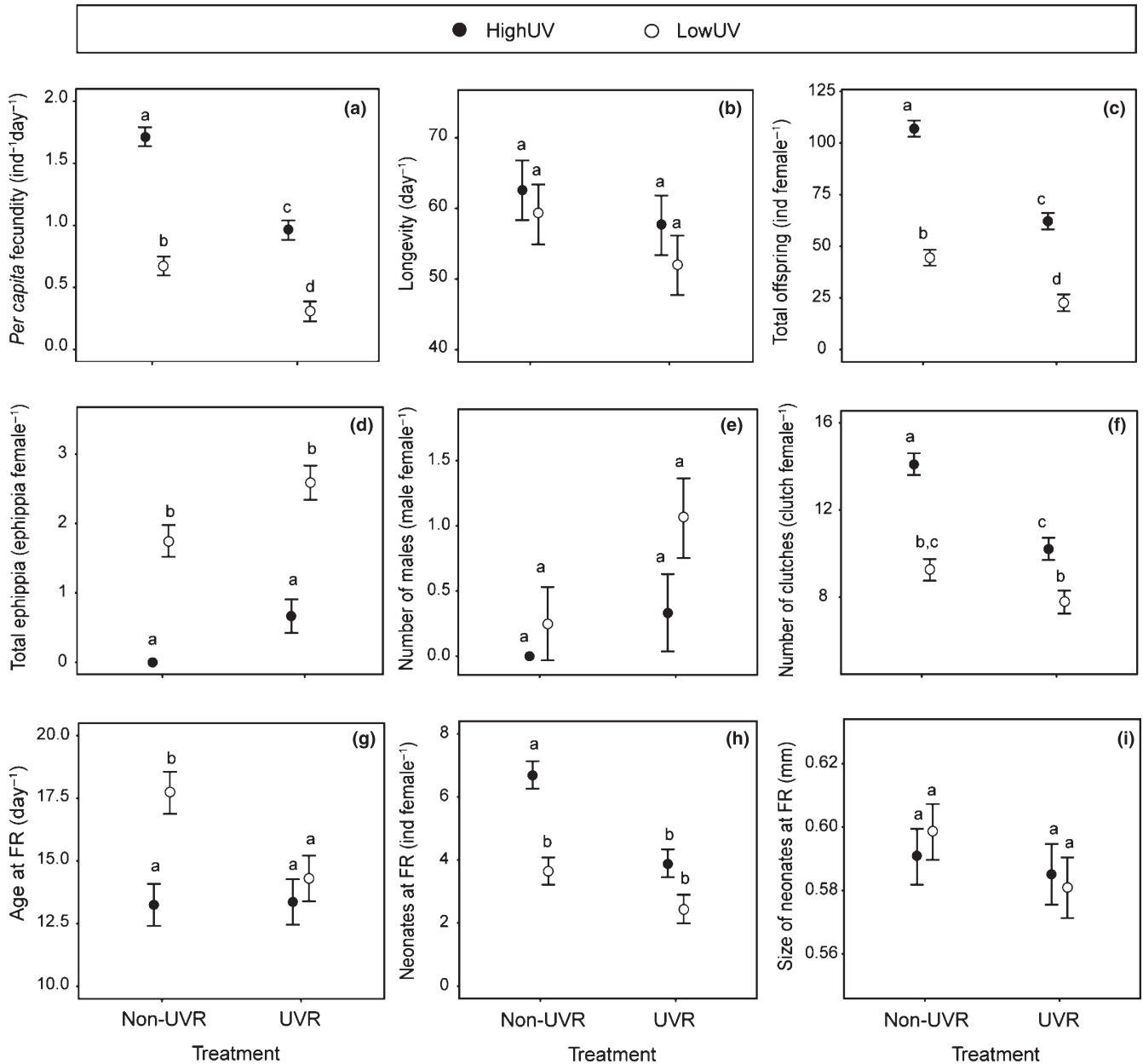

**FIGURE 4** Life-history responses as a function of *Daphnia pulex* lineages (dark circle = HighUV, open circle = LowUV) and UVR stress. (a) *per capita* fecundity, (b) Longevity, (c) total offspring per female, (d) total ephippia per female, (e) number of males per female (f) number of clutches per female, (g) age at first reproduction, (h) number of neonates at first reproduction, and (i) size of neonates at first reproduction. Means ± *SE* are shown. Letters above error bars represent statistically different groups (Tukey's post hoc test)

(Figure 4h). Although neonates tended to be smaller in the UVR treatment, there was no significant difference in offspring size between lineages or treatments (Figure 4i).

### 3.3 | Path analysis

#### 3.3.1 | Model fit

The obtained model for the HighUV and the LowUV life-history data had a comparative fit index (CFI) of 0.992 and a Tucker-Lewis index (TLI) of 0.986 suggesting a good fit. The model had a nonsignificant $\chi^2$ ($p = .30$) implying that the covariance structure specified by the model could not be rejected.

#### 3.3.2 | Total effects

A total effect represents how much of a given effect occurs due to a given shift in a precursor variable, regardless of the mechanisms by which the change may occur, and it is composed of partial effects that may be direct but also indirect (Alwin & Hauser, 1975). Considering total effects, UVR treatment negatively affected most of the life-history responses of *Daphnia* in both lineages. Almost all UVR

**TABLE 1** Results of two-way analyses of variances (ANOVA) of the mean life-history variables of HighUV and LowUV lineages of *Daphnia pulex*. All analyses compared the treatment (UVR and Non-UVR) to a lineage (HighUV and LowUV)

|  | df | SS | F | p |  | df | SS | F | p |
|---|---|---|---|---|---|---|---|---|---|
| **Per capita fecundity** |  |  |  |  | **Longevity** |  |  |  |  |
| Lineage | 1 | 13.85 | 119.22 | <.000*** | Lineage | 1 | 409.50 | 1.15 | .287 n.s. |
| Treatment | 1 | 6.07 | 52.23 | .000*** | Treatment | 1 | 738.10 | 2.07 | .154 n.s. |
| Treatment x lineage | 1 | 0.73 | 6.25 | .015* | Treatment x lineage | 1 | 25.30 | 0.07 | .791 n.s. |
| Residuals | 73 | 8.48 |  |  | Residuals | 71 | 27057.20 |  |  |
| **Total offspring** |  |  |  |  | **Total ephippia** |  |  |  |  |
| Lineage | 1 | 60044.00 | 201.98 | <.000*** | Lineage | 1 | 62.05 | 59.64 | .000*** |
| Treatment | 1 | 26272.00 | 88.38 | .000*** | Treatment | 1 | 10.53 | 10.12 | .002** |
| Longevity correction | 1 | 42282.00 | 142.23 | <.000*** | Treatment x lineage | 1 | 0.14 | 0.13 | .717 n.s. |
| Treatment x lineage | 1 | 2541.00 | 8.55 | .005* | Residuals | 71 | 73.87 |  |  |
| Residuals | 72 | 21404.00 |  |  |  |  |  |  |  |
| **Total males** |  |  |  |  | **Number of clutches** |  |  |  |  |
| Lineage | 1 | 4.03 | 2.54 | .115 n.s. | Lineage | 1 | 61.93 | 7.23 | .009** |
| Treatment | 1 | 6.01 | 3.79 | .056 n.s. | Treatment | 1 | 305.63 | 35.66 | .000*** |
| Treatment x lineage | 1 | 1.06 | 0.66 | .418 n.s. | Longevity correction | 1 | 57.54 | 6.71 | .012** |
| Residuals | 76 | 112.69 |  |  | Treatment x lineage | 1 | 26.98 | 3.15 | .080 n.s. |
|  |  |  |  |  | Residuals | 70 | 599.92 |  |  |
| **Age at first reproduction** |  |  |  |  | **Number of neonates at first reproduction** |  |  |  |  |
| Lineage | 1 | 135.14 | 9.58 | .003** | Lineage | 1 | 97.82 | 26.16 | .000*** |
| Treatment | 1 | 51.65 | 3.66 | .060 n.s. | Treatment | 1 | 78.15 | 20.90 | .000*** |
| Longevity correction | 1 | 41.36 | 2.93 | .091 n.s. | Treatment x lineage | 1 | 12.29 | 3.29 | .074 n.s. |
| Treatment x lineage | 1 | 57.18 | 4.05 | .048⁻ | Residuals | 73 | 272.98 |  |  |
| Residuals | 69 | 973.27 |  |  |  |  |  |  |  |
| **Size of neonates** |  |  |  |  |  |  |  |  |  |
| Lineage | 1 | 0.00 | 0.00 | .968 n.s. |  |  |  |  |  |
| Treatment | 1 | 0.09 | 10.43 | .002** |  |  |  |  |  |
| Longevity correction | 1 | 0.30 | 33.13 | .000*** |  |  |  |  |  |
| Treatment x lineage | 1 | 0.01 | 0.76 | .387 n.s. |  |  |  |  |  |
| Residuals | 72 | 0.65 |  |  |  |  |  |  |  |

Asterisks indicate significant differences (.000 "***", .001 "**", .01 "*", .05 "−", >.05 "n.s.").

treatment total effects were higher for the HighUV lineage, except for the number of offspring at FR, which was higher for the LowUV lineage.

### 3.3.3 | Partial effects

Reflects the decomposing of the total effects of the UVR treatment on all life-history responses into their constituent direct and indirect effects (Table 2, Figure 5). Four and five life-history variables in HighUV and LowUV lineages, respectively, showed direct effects. From them, total ephippia, total offspring, number of clutches, and number of offspring at FR were common between both lineages, while age at FR was directly affected by UVR only in the LowUV lineage model.

Ultraviolet radiation treatment had no direct effect on *per capita* fecundity on any of the lineages. Main indirect effects on *per capita* increase rate were those that affected offspring production.

Ultraviolet radiation had positive, direct, and indirect effects on total ephippia production of both lineages; the direct effect was the

TABLE 2 Total, direct, and indirect effects (standardized coefficients) of UVR treatment on each life-history responses of both lineages of *Daphnia pulex* in a structural equation model, as well as the mean (x̄) and standard deviation (SD) of each variable for Non-UVR and UVR treatments

| | HighUV | | | | | | LowUV | | | | | |
|---|---|---|---|---|---|---|---|---|---|---|---|---|
| | SEM | | | Non-UVR | UVR | | SEM | | | Non-UVR | UVR | |
| Life-history response | Total effect | Direct effect | Indirect effect | x̄ ± SD | x̄ ± SD | | Total effect | Direct effect | Indirect effect | x̄ ± SD | x̄ ± SD | |
| *Per capita* fecundity | −0.76 | 0 | −0.76 | 1.71 ± 0.38 | 0.96 ± 0.48 | | −0.35 | 0 | −0.35 | 0.67 ± 0.32 | 0.31 ± 0.17 | |
| Total ephippia | 0.58 | 0.07 | 0.51 | 0.00 ± 0.00 | 0.67 ± 0.75 | | 0.34 | 0.30 | 0.04 | 1.75 ± 1.37 | 2.59 ± 1.54 | |
| Longevity | −0.42 | 0 | −0.42 | 62.55 ± 16.59 | 57.60 ± 19.78 | | −0.23 | 0 | −0.23 | 59.15 ± 19.59 | 51.95 ± 19.33 | |
| Total offspring | −0.56 | −0.30 | −0.26 | 111.10 ± 37.32 | 62.74 ± 36.35 | | −0.46 | −0.20 | −0.26 | 43.65 ± 27.59 | 18.39 ± 10.20 | |
| Total number of males | 0 | 0 | 0 | 0.00 ± 0.00 | 0.33 ± 1.34 | | 0 | 0 | 0 | 0.25 ± 0.79 | 1.06 ± 1.92 | |
| Number of clutches | −0.47 | 0.38 | −0.09 | 14.85 ± 5.86 | 10.32 ± 5.13 | | −0.26 | −0.19 | −0.07 | 9.10 ± 5.45 | 7.00 ± 3.53 | |
| Age at FR | 0 | 0 | 0 | 13.35 ± 1.27 | 13.41 ± 5.44 | | −0.50 | −0.33 | −0.17 | 17.65 ± 5.49 | 14.24 ± 6.59 | |
| Size of offspring at FR | 0.45 | 0 | 0.45 | 0.59 ± 0.04 | 0.59 ± 0.22 | | 0.37 | 0 | 0.37 | 0.60 ± 0.04 | 0.58 ± 0.22 | |
| Number of offspring at FR | −0.55 | −0.55 | 0 | 6.70 ± 2.15 | 3.89 ± 2.34 | | −0.57 | −0.45 | −0.12 | 3.65 ± 1.66 | 2.44 ± 1.67 | |

FR, first reproduction; SEM, structural equations modeling; UVR, ultraviolet radiation.

most important for the LowUV lineage, while the indirect effects were the most important for the ephippia production of the HighUV lineage.

The number of males that were born during the experiment was not affected by the UVR treatment either directly or indirectly in any of the lineages. UVR had direct and indirect effects on the age at FR on the LowUV lineage but not on the HighUV lineage.

## 4 | DISCUSSION

Cladocerans have several efficient mechanisms to avoid UVR (Hansson et al., 2007; Rautio & Tartarotti, 2010). Here, we have demonstrated that their ability to exhibit different traits to counteract UVR damage is not only species-specific, but also differs between lineages within the same species but with different evolutionary histories, and that the involved trade-offs depend on the environment in which the organisms live.

Tolerance induction to UVR in terms of relaxed swimming behavior through exposition of several generations of *D. magna* has previously been studied (Hylander, Ekvall, Bianco, Yang, & Hansson, 2014). In this sense, an unanticipated finding of our study was that the *per capita* fecundity and almost all life-history responses of both accustomed (HighUV) and naive (LowUV) lineages were equally reduced, in percentage, under the UVR treatments (Figure 4). Some effects were even stronger in the HighUV than the LowUV lineage (total offspring, number of clutches, and number of offspring at FR), suggesting that, independently of their geographical origin, no higher physiological tolerance or plasticity has been induced by the HighUV lineage. Nevertheless, the elevated fecundity displayed by the HighUV lineage could easily withstand the losses imposed by the UVR stress. The most remarkable finding was that, regardless of the reduction that UVR caused in the *per capita* fecundity of the HighUV lineage, this was still higher than in the LowUV lineage, and the abundance of offspring produced by the HighUV lineage under UVR was superior to that produced by the LowUV lineage even without UVR (Figure 3).

As *Daphnia* have an extremely short generation time and most of Andean lakes were formed after the last deglaciation in the late Pleistocene (Seltzer, 1990), UVR as a selection force may have favored high-fertility lineages in such environments. It has been shown that artificial selection of high-fertility lineages may be achieved within a few generations (Langhammer et al., 2014). Mice, for example, require only about 160 generations to change their fertility capacity (Langhammer et al., 2014). Generation time is highly connected to age at first reproduction and may play an important evolutionary role (Kawecki & Ebert, 2004). In line with this, our LowUV lineage had its first reproduction (FR) at younger age under UVR exposure, which is a typical phenotypic plasticity response to new environmental conditions (Barata, Baird, & Soares, 2001; Fischer & Fiedler, 2002), to a strong stressor (Spitze, 1991), or to a seasonal trigger (Nylin, 1992; Nylin & Gotthard, 1998). On the other hand, the HighUV lineage started to reproduce earlier even without UVR exposure, suggesting that this lineage is either adapted to higher levels of UVR stress or that its trigger to reproduce is not associated with UVR. These differences may have been influenced by historic environmental UVR seasonality; zooplankton in Andean lakes suffer high UV stress year around, whereas animals in Swedish lakes are exposed to moderate or low UVR stress in spring-summer, and fall-winter, respectively. Hence, whereas the LowUV lineage animals were forced to respond with phenotypic plasticity to the elevated experimental UVR conditions, the HighUV lineage animals required no warning trigger to start reproducing earlier, as they were already adapted to intense UVR conditions. These results seem to be consistent with previous research showing that high fecundity and early age at FR differentiate Antarctic tardigrades from their temperate relatives, suggesting that these two characteristics constitute a strategy to cope with extreme and stochastic environmental conditions (Altiero, Giovannini, Guidetti, & Rebecchi, 2015). In this sense, we have here disentangled an alternative strategy that does not prevent or limit the UVR damage, but that handle UVR effects by maintaining fitness and a stable population size through high fecundity and early reproduction.

In our study, the longevity was not significantly or directly affected by the UVR treatment for any of the lineages. In contrast to the reduced survival reported as a consequence of UV-B stress on cladoceran species (Connelly et al., 2015; Huebner, Loadman, Wiegand, Young, & Warszycki, 2009), low doses of UV-A radiation may not significantly affect the life span of *Daphnia*, although reproduction could be affected (Zellmer, 1998). In this case, the structural equation model analysis revealed that all effects of UVR on the *per capita* fecundity were likely indirect effects for both lineages (Figure 5, Table 2). The *per capita* fecundity of both lineages was affected by the numbers of ephippia, offspring, and clutches, respectively. Nevertheless, there was a clear difference between structures of the models of both lineages (Figure 5): while both lineages were negatively affected regarding age/number of offspring at FR and positively influenced regarding size of offspring at FR, this trade-off (fewer but bigger offspring, with better chances to survive; Dudycha & Tessier, 1999) had no influence on the *per capita* fecundity of the HighUV lineage.

The total lack of production of ephippia in the HighUV lineage in the absence of UVR (Table 2) suggests that under certain circumstances, some lineages of *D. pulex* do not invest in ephippia production (at least within the time frame and environmental conditions of our experimental setup), thereby increasing the number of clutches and the number of offspring. Interestingly, at high UVR

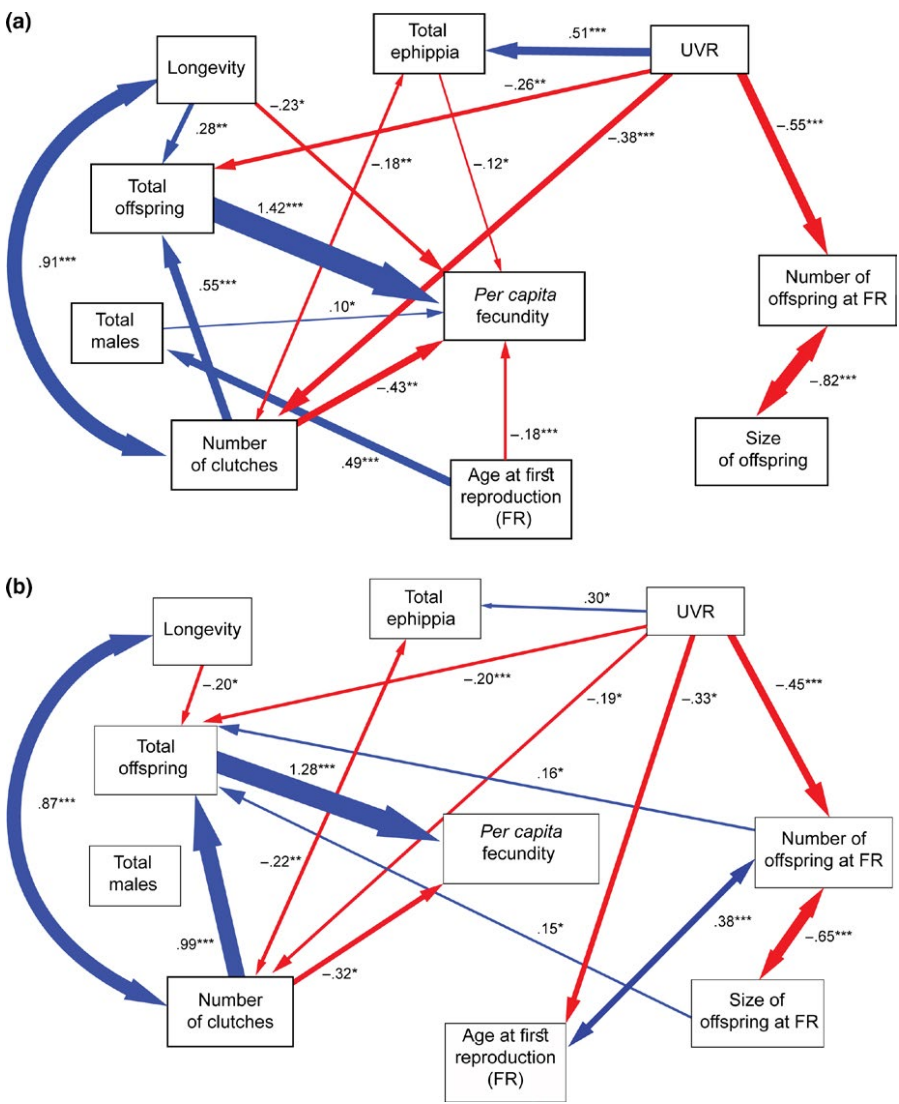

**FIGURE 5** Structural equation model of the influence of UVR on life-history responses of *Daphnia pulex*. Resulting path models for: (a) HighUV and (b) LowUV lineage. Arrow widths are proportional to path coefficients. One-headed arrows depict causal relationships whereas two-headed arrows depict correlations. The sign of coefficients denotes positive or negative effects. The level of statistical significance is indicated by asterisks (*$p < .05$; **$p < .01$; ***$p < .001$)

stress, animals initiated the production of ephippia, allowing them to secure the maintenance of the population in the future. UVR treatment lead to reduced number of clutches and increased number of ephippia in both lineages, which may reflect a trade-off between investing energy in the present or in the future generations, supporting earlier observations of lower proportion of egg-carrying females and higher production of ephippia as a consequence of UVR exposure (Hylander & Hansson, 2010).

In conclusion, by disentangling direct and indirect effects of UVR on both high and low UVR adapted lineages, we show here that UVR affects almost all life-history variables of both lineages, although they have adopted different life-history traits to cope with UVR. Moreover, in contrast to our initial hypothesis that High UVR adapted lineages should be more tolerant to UVR, both lineages had relatively similar ability to handle UVR stress. Instead, our results suggest that UVR might act as a selection force and increased fecundity and earlier age at first reproduction could be a major evolutionary strategy to manage the losses caused by UVR. However, our experiment standardized environmental factors, including water chemistry, temperature, photoperiod and food quality/quantity, setting aside seasonality, and predation pressure. Hence, further studies are needed to understand how multiple environmental factors have molded *Daphnia* life-history strategies to succeed in a specific habitat.

## ACKNOWLEDGMENTS

We would like to thank Mikael Ekvall for helping in logistics and technical issues in the experiment and two anonymous referees for their valuable advice. This work was financially supported by a research grant from the Swedish International Development Cooperation Agency (SIDA), GIRH project, and the Swedish Research Council (VR to LAH).

## CONFLICT OF INTEREST

None declared.

## AUTHOR CONTRIBUTIONS

CF, MC, and L-AH conceived the ideas and designed the experiment; CF and L-AH collected the data; CU led the revision of the methods; CF and MC led the analysis of data; CF wrote the initial version of the manuscript. All authors contributed critically to the drafts and gave final approval for publication.

## ORCID

*Carla E. Fernández*  http://orcid.org/0000-0002-2200-5991

*Melina Campero*  http://orcid.org/0000-0001-7878-4261

*Cintia Uvo*  http://orcid.org/0000-0002-8497-0295

*Lars-Anders Hansson*  http://orcid.org/0000-0002-3035-1317

## REFERENCES

Aguilera, X., Lazzaro, X., & Coronel, J. S. (2013). Tropical high-altitude Andean lakes located above the tree line attenuate UV-A radiation more strongly than typical temperate alpine lakes. *Photochemical & Photobiological Sciences*, *12*, 1649. https://doi.org/10.1039/c3pp25285j

Alpert, P., & Simms, E. L. (2002). The relative advantages of plasticity and fixity in different environments: When is it good for a plant to adjust? *Evolutionary Ecology*, *16*, 285–297.

Altiero, T., Giovannini, I., Guidetti, R., & Rebecchi, L. (2015). Life history traits and reproductive mode of the tardigrade *Acutuncus antarcticus* under laboratory conditions: Strategies to colonize the Antarctic environment. *Hydrobiologia*, *761*, 277–291. https://doi.org/10.1007/s10750-015-2315-0

Alwin, D. F., & Hauser, R. M. (1975). The decomposition of effects in path analysis. *American Sociological Review*, *40*, 37–47. https://doi.org/10.2307/2094445

Barata, C., Baird, D., & Soares, A. (2001). Phenotypic plasticity in *Daphnia magna* Straus: Variable maturation instar as an adaptive response to predation pressure. *Oecologia*, *129*, 220–227. https://doi.org/10.1007/s004420100712

Bentler, P. M. (1990). Comparative fit indexes in structural models. *Psychological Bulletin*, *107*, 238. https://doi.org/10.1037/0033-2909.107.2.238

Bornman, J. F., Barnes, P. W., Robinson, S. A., Ballare, C. L., Flint, S. D., & Caldwell, M. M. (2015). Solar ultraviolet radiation and ozone depletion-driven climate change: Effects on terrestrial ecosystems. *Photochemical & Photobiological Sciences*, *14*(1), 88–107. https://doi.org/10.1039/C4PP90034K

Brommer, J. E., Gustafsson, L., Pietiäinen, H., & Merilä, J. (2004). Single-generation estimates of individual fitness as proxies for long-term genetic contribution. *The American Naturalist*, *163*, 505–517. https://doi.org/10.1086/382547

Callahan, H. S., Dhanoolal, N., & Ungerer, M. C. (2005). Plasticity genes and plasticity costs: A new approach using an *Arabidopsis* recombinant inbred population. *New Phytologist*, *166*, 129–139. https://doi.org/10.1111/j.1469-8137.2005.01368.x

Campero, M., Moreira, R., Lucano, M., & Rejas, D. (2011). Coeficientes de atenuación y profundidad de penetración de la radiación ultravioleta en 14 lagunas altoandinas de Bolivia. *Revista Boliviana de Ecología y Conservación Ambiental*, *29*, 83–86.

Connelly, S. J., Walling, K., Wilbert, S. A., Catlin, D. M., Monaghan, C. E., Hlynchuk, S., … Bowles, S. M. (2015). UV-stressed *Daphnia pulex* increase fitness through uptake of Vitamin D3. *PLoS One*, *10*(7), e0131847. https://doi.org/10.1371/journal.pone.0131847

Danilov, R. A., & Ekelund, N. G. (2001). Effects of solar radiation, humic substances and nutrients on phytoplankton biomass and distribution in Lake Solumsjö, Sweden. *Hydrobiologia*, *444*, 203–212. https://doi.org/10.1023/A:1017535830980

Dudycha, J. L., & Tessier, A. J. (1999). Natural genetic variation of life span, reproduction, and juvenile growth in *Daphnia*. *Evolution*, *53*, 1744–1756. https://doi.org/10.1111/j.1558-5646.1999.tb04559.x

Ekvall, M. T., Hylander, S., Walles, T., Yang, X., & Hansson, L.-A. (2015). Diel vertical migration, size distribution and photoprotection in zooplankton as response to UV-A radiation. *Limnology and Oceanography*, *60*, 2048–2058. https://doi.org/10.1002/lno.10151

Erickson, D. J. III, Sulzberger, B., Zepp, R. G., & Austin, A. T. (2015). Effects of stratospheric ozone depletion, solar UV radiation, and climate change on biogeochemical cycling: Interactions and feedbacks. *Photochemical & Photobiological Sciences*, *14*(1), 127–148. https://doi.org/10.1039/C4PP90036G

Fernández, C. E., & Rejas, D. (2017). Effects of UVB radiation on grazing of two cladocerans from high-altitude Andean lakes. *PLoS One*, *12*, e0174334. https://doi.org/10.1371/journal.pone.0174334

Finkel, Z. V., Beardall, J., Flynn, K. J., Quigg, A., Rees, T. A. V., & Raven, J. A. (2009). Phytoplankton in a changing world: Cell size and elemental stoichiometry. *Journal of Plankton Research*, *32*(1), 119–137.

Fischer, K., & Fiedler, K. (2002). Reaction norms for age and size at maturity in response to temperature: A test of the compound interest hypothesis. *Evolutionary Ecology*, *16*, 333–349. https://doi.org/10.1023/A:1020271600025

Gabriel, W., Luttbeg, B., Sih, A., & Tollrian, R. (2005). Environmental tolerance, heterogeneity, and the evolution of reversible plastic responses. *The American Naturalist*, *166*, 339–353. https://doi.org/10.1086/432558

Häder, D. P., Helbling, E. W., Williamson, C. E., & Worrest, R. C. (2011). Effects of UV radiation on aquatic ecosystems and interactions with climate change. *Photochemical & Photobiological Sciences*, *10*, 242–260. https://doi.org/10.1039/c0pp90036b

Häder, D. P., Williamson, C. E., Wangberg, S. A., Rautio, M., Rose, K. C., Gao, K., … Worrest, R. (2015). Effects of UV radiation on aquatic ecosystems and interactions with other environmental factors. *Photochemical & Photobiological Sciences*, *14*, 108–126. https://doi.org/10.1039/C4PP90035A

Hansson, L.-A. (2004). Plasticity in pigmentation induced by conflicting threats from predation and UV radiation. *Ecology*, *85*, 1005–1016. https://doi.org/10.1890/02-0525

Hansson, L.-A., & Hylander, S. (2009). Effects of ultraviolet radiation on pigmentation, photoenzymatic repair, behavior, and community ecology of zooplankton. *Photochemical & Photobiological Sciences*, *8*, 1266–1275. https://doi.org/10.1039/b908825c

Hansson, L.-A., Hylander, S., & Sommaruga, R. (2007). Escape from UV threats in zooplankton: A cocktail of behavior and protective pigmentation. *Ecology*, *88*, 1932–1939. https://doi.org/10.1890/06-2038.1

Hays, G. C., Richardson, A. J., & Robinson, C. (2005). Climate change and marine plankton. *Trends in Ecology & Evolution*, *20*, 337–344. https://doi.org/10.1016/j.tree.2005.03.004

Huebner, J. D., Loadman, N. L., Wiegand, M. D., Young, D. L., & Warszycki, L. A. (2009). The effect of chronic exposure to artificial UVB radiation on the survival and reproduction of *Daphnia magna* across two generations. *Photochemistry and Photobiology*, *85*, 374–378. https://doi.org/10.1111/j.1751-1097.2008.00454.x

Hylander, S., Ekvall, M. T., Bianco, G., Yang, X., & Hansson, L. A. (2014). Induced tolerance expressed as relaxed behavioural threat response in millimetre-sized aquatic organisms. *Proceedings of the Royal Society of London. Series B, Biological Sciences*, *281*, 20140364. https://doi.org/10.1098/rspb.2014.0364

Hylander, S., & Hansson, L. A. (2010). Vertical migration mitigates UV effects on zooplankton community composition. *Journal of Plankton Research*, *32*, 971–980. https://doi.org/10.1093/plankt/fbq037

Kawecki, T. J., & Ebert, D. (2004). Conceptual issues in local adaptation. *Ecology Letters*, *7*, 1225–1241. https://doi.org/10.1111/j.1461-0248.2004.00684.x

Kessler, K., Lockwood, R. S., Williamson, C. E., & Saros, J. E. (2008). Vertical distribution of zooplankton in subalpine and alpine lakes: Ultraviolet radiation, fish predation, and the transparency-gradient hypothesis. *Limnology and Oceanography*, *53*, 2374–2382. https://doi.org/10.4319/lo.2008.53.6.2374

Lampert, W. (1993). Phenotypic plasticity of the size at 1st reproduction in *Daphnia* - the importance of maternal size. *Ecology*, *74*, 1455–1466. https://doi.org/10.2307/1940074

Langhammer, M., Michaelis, M., Hoeflich, A., Sobczak, A., Schoen, J., & Weitzel, J. M. (2014). High-fertility phenotypes: Two outbred mouse models exhibit substantially different molecular and physiological strategies warranting improved fertility. *Reproduction*, *147*, 427–433. https://doi.org/10.1530/REP-13-0425

Leroi, A. M., Chippindale, A. K., & Rose, M. R. (1994). Long-term laboratory evolution of a genetic life-history trade-off in *Drosophila melanogaster*. 1. The role of genotype-by-environment interaction. *Evolution*, *48*, 1244–1257.

Lynch, M. (1983). Ecological genetics of *Daphnia pulex*. *Evolution*, *37*, 358–374. https://doi.org/10.1111/j.1558-5646.1983.tb05545.x

Medina-Sánchez, J. M., Delgado-Molina, J. A., Bratbak, G., Bullejos, F. J., Villar-Argaiz, M., & Carrillo, P. (2013). Maximum in the middle: Nonlinear response of microbial plankton to ultraviolet radiation and phosphorus. *PLoS One*, *8*(4), e60223. https://doi.org/10.1371/journal.pone.0060223

Mitchell, D. L., & Karentz, D. (1993). The induction and repair of DNA photodamage in the environment. In A. R. Young, L. O. Bjorn, J. Moan, & W. Nultsch (Eds.), *Environmental UV photobiology* (pp. 345–377). New York, NY: Springer.

Mitchell, S., & Lampert, W. (2000). Temperature adaptation in a geographically widespread zooplankter, *Daphnia magna*. *Journal of Evolutionary Biology*, *13*, 371–382. https://doi.org/10.1046/j.1420-9101.2000.00193.x

Morgan-Kiss, R. M., Priscu, J. C., Pocock, T., Gudynaite-Savitch, L., & Huner, N. P. (2006). Adaptation and acclimation of photosynthetic microorganisms to permanently cold environments. *Microbiology and Molecular Biology Reviews*, *70*(1), 222–252. https://doi.org/10.1128/MMBR.70.1.222-252.2006

Nylin, S. (1992). Seasonal plasticity in life history traits: Growth and development in Polygonia c-album (Lepidoptera: Nymphalidae). *Biological Journal of the Linnean Society*, *47*, 301–323. https://doi.org/10.1111/j.1095-8312.1992.tb00672.x

Nylin, S., & Gotthard, K. (1998). Plasticity in life-history traits. *Annual Review of Entomology*, *43*, 63–83. https://doi.org/10.1146/annurev.ento.43.1.63

Pugesek, B. H., Tomer, A., & Von Eye, A. (2003). *Structural equation modeling: Applications in ecological and evolutionary biology*. Cambridge, UK: Cambridge University Press. https://doi.org/10.1017/CBO9780511542138

PV Lighthouse Pty (2014). *Solar spectrum simulation tool*, Version 1.1.1. PV Lighthouse Pty. Ltd., Australia. Retrieved from http://www.pvlighthouse.com.au

R Core Team (2016). *R: A language and environment for statistical computing*. Vienna, Austria: R Foundation for Statistical Computing. Retrieved from https://www.R-project.org/

Rautio, M., Bonilla, S., & Vincent, W. F. (2009). UV photoprotectants in arctic zooplankton. *Aquatic Biology*, *7*, 93–105. https://doi.org/10.3354/ab00184

Rautio, M., & Korhola, A. (2002). UV-induced pigmentation in subarctic *Daphnia*. *Limnology and Oceanography*, *47*, 295–299. https://doi.org/10.4319/lo.2002.47.1.0295

Rautio, M., & Tartarotti, B. (2010). UV radiation and freshwater zooplankton: Damage, protection and recovery. *Freshwater Reviews*, *3*, 105–131. https://doi.org/10.1608/FRJ-3.2.157

Rosseel, Y. (2012). *Lavaan: An R package for structural equation modeling and more. Version 0.5–12 (BETA)*. Ghent, Belgium: Ghent University.

Seltzer, G. O. (1990). Recent glacial history and paleoclimate of the Peruvian-Bolivian Andes. *Quaternary Science Reviews*, *9*, 137–152. https://doi.org/10.1016/0277-3791(90)90015-3

Sommaruga, R. (2003). Ultraviolet radiation and its effects on species interactions. In E. W. Helbling, & H. Zagarese (Eds.), *UV effects in aquatic organisms and ecosystems*. London, UK: European Society for Photobiology, Royal Society of Chemistry.

Sommaruga, R., & Buma, A. G. J. (2000). UV-induced cell damage is species-specific among aquatic phagotrophic protists. *Journal of Eukaryotic Microbiology*, *47*, 450–455. https://doi.org/10.1111/j.1550-7408.2000.tb00074.x

Spitze, K. (1991). *Chaoborus* predation and life-history evolution in *Daphnia pulex*: Temporal pattern of population diversity, fitness, and mean life history. *Evolution*, *45*, 82–92.

Taghavi, D., Farhadian, O., Soofiani, N. M., & Keivany, Y. (2013). Effects of different light/dark regimes and algal food on growth, fecundity, ephippial induction and molting of freshwater cladoceran, *Ceriodaphnia*

*quadrangula*. *Aquaculture*, *410*, 190–196. https://doi.org/10.1016/j.aquaculture.2013.06.026

Tollrian, R., & Heibl, C. (2004). Phenotypic plasticity in pigmentation in *Daphnia* induced by UV radiation and fish kairomones. *Functional Ecology*, *18*, 497–502. https://doi.org/10.1111/j.0269-8463.2004.00870.x

Tucker, L. R., & Lewis, C. (1973). A reliability coefficient for maximum likelihood factor analysis. *Psychometrika*, *38*, 1–10. https://doi.org/10.1007/BF02291170

Villafañe, V. E., Andrade, M., Lairanat, V., Zaratti, F., & Helbling, E. W. (1999). Inhibition of phytoplankton photosynthesis by solar ultraviolet radiation: Studies in Lake Titicaca, Bolivia. *Freshwater Biology*, *42*, 215–224. https://doi.org/10.1046/j.1365-2427.1999.444453.x

Wickham, H. (2009). *Elegant graphics for data analysis (ggplot2)*. New York, NY: Springer-Verlag.

Williamson, C. E., Zepp, R. G., Lucas, R. M., Madronich, S., Austin, A. T., Ballaré, C. L., … Robinson, S. A. (2014). Solar ultraviolet radiation in a changing climate. *Nature Climate Change*, *4*(6), 434–441. https://doi.org/10.1038/nclimate2225

Winder, M., & Schindler, D. E. (2004). Climate change uncouples trophic interactions in an aquatic ecosystem. *Ecology*, *85*, 2100–2106. https://doi.org/10.1890/04-0151

Zagarese, H. E., Feldman, M., & Williamson, C. E. (1997). UV-B-induced damage and photoreactivation in three species of *Boeckella* (Copepoda, Calanoida). *Journal of Plankton Research*, *19*, 357–367. https://doi.org/10.1093/plankt/19.3.357

Zellmer, I. D. (1998). The effect of solar UVA and UVB on subarctic *Daphnia pulicaria* in its natural habitat. *Hydrobiologia*, *379*, 55–62. https://doi.org/10.1023/A:1003285412043